\documentclass[citeautoscript,floatfix,aps,prb,twocolumn,superscriptaddress]{revtex4-2}
\usepackage{graphicx}
\usepackage{dcolumn}
\usepackage{bm}

\usepackage{nicefrac}
\usepackage{hyperref}
\usepackage{booktabs}
\usepackage{siunitx}
\usepackage{import}
\usepackage{makecell}
\usepackage[caption=false]{subfig}

\def \questaal{\textit{Questaal }} 
\def \bse{QSG$\widehat{W}$ } 

\hypersetup{
colorlinks=true,
linkcolor=blue,
filecolor=magenta,
urlcolor=blue,
citecolor= blue,
bookmarks=true,
}

\usepackage{graphicx}
\usepackage{amsmath}
\usepackage{amssymb}
\usepackage{dcolumn}
\usepackage[version=4]{mhchem}
\usepackage{latexsym}
\usepackage{rotating}
\usepackage{epstopdf}
\usepackage[usenames,dvipsnames]{xcolor}
\usepackage{float}
\usepackage{soul}
\usepackage{epsfig}
\usepackage{psfrag}
\usepackage{natbib}
\usepackage{bm}
\usepackage{eucal}
\usepackage{mathrsfs}
\usepackage{braket}
\usepackage{enumerate}
\usepackage{longtable}
\usepackage{bm}
\usepackage{hyperref}
\usepackage{amsfonts}
\usepackage{dcolumn}
\usepackage{bm}
\usepackage{changes}

\newcommand{\be}{\begin{equation}}
	\newcommand{\ee}{\end{equation}}
\newcommand{\bn}{\begin{eqnarray}}
	\newcommand{\en}{\end{eqnarray}}

\def\x2y2{{x^2-y^2}}

\usepackage{color} 

\usepackage{hyperref}
\hypersetup{
	colorlinks=true,final=true,
	linkcolor=blue,
	citecolor=blue,
	filecolor=blue,
	urlcolor=blue,
}

\begin{document}


\title{Possible realization of hyperbolic plasmons in a few-layered rhenium disulfide}
\author{Ravi Kiran}
\email{ravieroy123@iitkgp.ac.in}
\affiliation{Indian Institute of Technology Kharagpur, Kharagpur 721302, West Bengal, India} 
\author{Dimitar Pashov}
\affiliation{ King's College London, Theory and Simulation of Condensed Matter,
	The Strand, WC2R 2LS London, UK}       
\author{Mark van Schilfgaarde}
\affiliation{National Renewable Energy Laboratories, Golden, CO 80401, USA} 
\affiliation{ King's College London, Theory and Simulation of Condensed Matter,
	The Strand, WC2R 2LS London, UK}
\author{Mikhail I. Katsnelson}        
\affiliation{Institute for Molecules and Materials, Radboud University, NL-6525 AJ Nijmegen, The Netherlands}
\author{A. Taraphder}
\affiliation{Indian Institute of Technology Kharagpur, Kharagpur 721302, West Bengal, India}
\author{Swagata Acharya} 
\email{swagata.acharya@nrel.gov}
\affiliation{National Renewable Energy Laboratories, Golden, CO 80401, USA}     
\affiliation{Institute for Molecules and Materials, Radboud University, NL-6525 AJ Nijmegen, The Netherlands}


\begin{abstract}
The in-plane structural anisotropy in low-symmetric layered compound rhenium disulfide ($\text{ReS}_2$) makes it a candidate to host and tune electromagnetic phenomena specific for anisotropic media. In particular, optical anisotropy may lead to the appearance of hyperbolic plasmons, a highly desired property in optoelectronics. The necessary condition is a strong anisotropy of the principal components of the dielectric function, such that at some frequency range, one component is negative and the other is positive, i.e., one component is metallic, and the other one is dielectric. Here, we study the effect of anisotropy in $\text{ReS}_2$ and show that it can be a natural material to host hyperbolic plasmons in the ultraviolet frequency range. The operating frequency range of the hyperbolic plasmons can be tuned with the number of $\text{ReS}_2$ layers. 
\end{abstract}

\maketitle



\section{\textsc{Introduction}}

The rise of hyperbolic materials in recent years promises important applications in
optoelectronics and nanophotonics~\cite{hyperbolic1,hyperbolic2,hyperbolic3,hyperbolic4,CuS19}. Light can acquire
hyperbolic dispersion while passing through such materials, which occurs in some frequency range, when different principal components of the longitudinal dielectric function (dielectric permittivity) have opposite signs. For the case of isotropic medium, it behaves as a dielectric, that is, supports propagating electromagnetic waves, when the sign of the dielectric function is positive. When the latter is negative, the incident light is reflected, with only an exponentially decaying evanescent field penetrating the material, like for metals below plasma threshold. 

Anisotropy in electronic, optical, vibrational, and transport behaviour can occur when structural anisotropy is present, and if it is sufficiently strong, the different components of the dielectric permittivity tensor may aquire opposite signs to turn the material hyperbolic.  In two-dimensional crystals, in-plane anisotropy strong enough to make it hyperbolic is a unique situation that allows to confine short wavelengths (large wave vectors) inside a material, promising smaller sizes for optoelectronic devices.  Optical anisotropy in rhenium disulfide ($\text{ReS}_2$) has been established both for bulk crystals \cite{lin2011anisotropy,friemelt1993optical} and for thin layers \cite{liu2015integrated}.  In this work we predict that ($\text{ReS}_2$) that appears in a distorted 1T phase can realize hyperbolic plasmons depending on the number of layers.

It has been suggested that anisotropic 2D materials can be tuned to become hyperbolic via electrostatic tuning, strain or dimensionality and can host hyperbolic plasmons \cite{nemilentsau2016anisotropic,van2019tuning}. Several studies \cite{low2017polaritons,sun2014indefinite,cordova2019anisotropic} have investigated hyperbolic plasmons (HP) and its existence in naturally occurring materials. The strong anisotropy of $\text{ReS}_2$ hints at its potential as a natural hyperbolic material, offering possibilities for studying HP. Previous works
\cite{echeverry2018theoretical,zhang2015res2} have studied the band structure and anisotropic optical response of
$\text{ReS}_2$, but the study of HP remains largely unexplored. Here we use the ladder-vertex corrected and local-field corrected plasmonic response in ReS$_{2}$ within a self-consistent solution of Bethe-Salpeter equation (BSE) as implemented in \questaal \cite{pashov2020questaal}.

%
%

The rest of the paper is organized as follows. In Sec. \ref{sec: atomic-str}, we describe the anisotropic atomic structure of $\text{ReS}_2$ and in Sec. \ref{sec: comp-details}, we briefly describe theoretical methods and provide computational details. In Secs. \ref{sec: band-str} and \ref{sec: optical-spectra} we present our
results on the electronic structure and optical properties of $\text{ReS}_2$. In Sec. \ref{sec: conclusion}, we briefly summarize our results and conclude the paper.


\section{\textsc{Atomic Structure and Computational Details}}
\subsection{\textsc{Atomic structure}} \label{sec: atomic-str}

\begin{figure*}[htb]
	\includegraphics[scale=0.3]{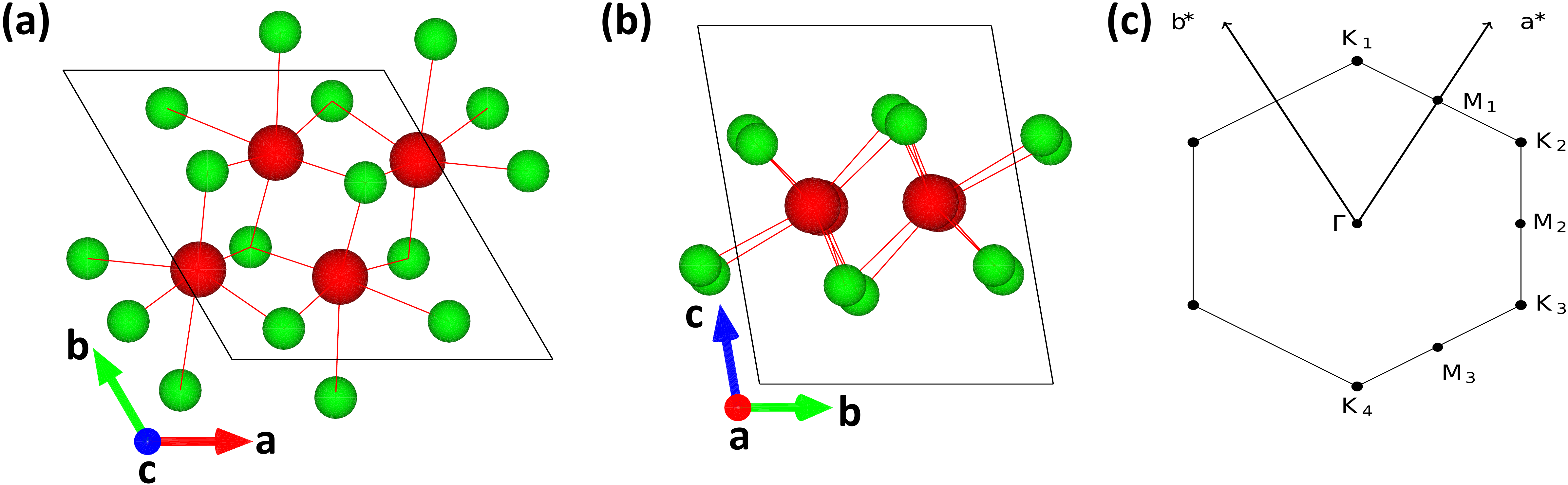}%
	\caption{The ball-stick model of the bulk distorted 1T diamond-chain $\text{ReS}_2$ obtained using VESTA software \cite{momma2011vesta} is presented. Top view (Fig.1(a)) and side view (Fig.1(b))of the distorted 1T-$\text{ReS}_2$; Re atoms are in red and S atoms are in green. The black outline shows the unit cell used for the calculation. The Re chain is along b direction. Brillouin zone (bottom) of the corresponding hexagonal lattice with lines connecting high-symmetry points $\Gamma$-K1-K2-M2-$\Gamma$-K4-K3-M2-$\Gamma$. $a^*$ and $b^*$ denote reciprocal
		lattice vectors.}
	\label{fig:res2-str}
	
\end{figure*}
%
%
%
	
$\text{ReS}_2$ belongs to the family of two-dimensional (2D) layered transition metal dichalcogenide (TMDs) of the form $\text{MX}_2$ where M is a transition metal atom (Mo, W, Re, ...) and X is a group-16 atom (S, Se, Te). The atomic structure of $\text{ReS}_2$ layers has neither H or T character. Unlike other TMDs, which usually have 1H or 1T structure in their ground state, $\text{ReS}_2$ crystallizes in a distorted-1T structure with clustering of Re units forming parallel metal chains along the van der Waals plane (see Fig. \ref{fig:res2-str}).
	
The compound $\text{ReS}_2$ belongs to the triclinic symmetry group $\text{P}^1$, resembling a distorted CdCl$_{2}$ structure. It comprises three atomic layers, S-Re-S, where covalent bonds join Re and S. The adjacent layers of $\text{ReS}_2$ are coupled by weak van der Waals (vdW) forces to form bulk crystals. The unit cell is derived from hexagonal symmetry towards a distorted 1T structure, in which Re atoms group into parallelograms of four Re atoms. The formation of Re chains breaks the hexagonal symmetry and doubles the unit cell size.  Hence the unit cell of single layer $\text{ReS}_2$ in the distorted-1T phase is composed of four Re and eight S atoms. In pristine $\text{ReS}_2$, the valence band maximum is composed from 5d orbitals of Re atoms and 3p orbitals of S atoms, and the conduction band minimum is derived from 5d orbitals of Re atoms. The Brillouin zone of the $\text{ReS}_2$ is hexagonal but with unequal sides as a result of the distorted atomic structure.  Energy band structures were generated along the symmetry lines shown in Fig. \ref{fig:res2-str}.

\subsection{\textsc{Computational Details} } \label{sec: Computational_details}\label{sec: comp-details}

\subsubsection{\textbf{LDA, QS\emph{GW}, and \bse self-consistency}}

Single-particle calculations (LDA, and the the quasi-particle self-consistent \emph{GW}~\cite{van2006quasiparticle}
(QS\emph{GW}) self-energy $\Sigma^0(k)$) were performed on a $12 \times 12 \times 12$ points (Monkhorst pack) for bulk
and $12 \times 12 \times 1$ for ML and BL. An energy cutoff of 400 \si{\electronvolt} was used and Gaussian smearing
with a width of 0.05 \si{\electronvolt}. The tolerance of $10^{-5}$ \si{\electronvolt} and $2 \times 10^{-6}$ has been
taken in convergence of energy and RMS density respectively. The charge density was made self-consistent for each
iteration in the QS\emph{GW} self-consistency cycle. The QS\emph{GW} cycle was iterated until the RMS change in
$\Sigma^0$ reached $10^{-5}$ Ry. Thus the calculation was self-consistent in both $\Sigma^0(k)$ and the
density. Numerous checks were made to verify that the self-consistent $\Sigma^0(k)$(k) was independent of starting
point.  For ML-$\text{ReS}_2$, we performed a rigorous check for vacuum correction to all band gap and dielectric
screening by increasing the size from \SI{10}{\angstrom} to \SI{45}{\angstrom}. Since along the \emph{z}-direction we
have a vacuum, the dielectric constant, which is the real part of the macroscopic dielectric response at $\omega{=}0$,
should be close to unity.
	
In the present work, the electron-hole two-particle correlations are incorporated within a
self-consistent ladder BSE implementation \cite{cunningham2018effect} with Tamn-Dancoff approximation
\cite{hirata1999time}. Ladder diagrams are included in the polarizability $P$ that makes $W$, via the
  solution of a Bethe-Salpeter equation (BSE); thus this form of $GW$ goes beyond the RPA in constructing the
  self-energy $\Sigma=iGW$. The electron-hole attraction from the ladders enhances $P$, thus reducing $W$, which in turn
  reduces the bandgap.  A static vertex is used to construct $P$.  \emph{G} and \emph{W} are calculated
  self-consistently, in quasiparticlized form~\cite{van2006quasiparticle}: \emph{G} and \emph{W} are updated iteratively
  until all of them converge (QS\emph{GW}).  When ladders are incorporated into \emph{W}, we denote the process as
  QS$G\widehat{W}$ to signify $W$ was computed from the BSE.  The macroscopic dielectric function we present here,
  $[\epsilon^{-1}_{\mathbf{G}{=}0,\mathbf{G'}{=}0}(q{\rightarrow}0,\omega)]^{-1}$, is also computed with the BSE.

The tetrahedron method is employed for integration over the Brillouin zone to calculate the optical spectrum. When
calculating the dielectric response within BSE, the valence and conduction states that form the two-particle Hamiltonian
are increased until the two-particle eigenvalues converge within an accuracy of 10 meV. For ReS$_{2}$ the excitons are
essentially Wannier-Mott~\cite{qiu2013optical} in nature and only the states at the valence band top and conduction band
bottom contribute to their formation, so the convergence in the two-particle Hamiltonian size is much faster compared to
the cases of CrX$_{3}$~\cite{acharya2022real} where the excitons have Frenkel character and many valence and conductions
bands over several electron volts form them. However, for the present work, our focus is the plasmonic response, while
we will note the excitonic binding energies in different layered variants later.
	
Table (\ref{tab:Lattice_Parameters}) contains the lattice parameters for $\text{ReS}_2$ used throughout the
calculations.
	
%

\begin{table*}[!ht]
	\centering
	\caption{Lattice parameters of bulk, monolayer (ML) and bilayer (BL) $\text{ReS}_2$}
	\label{tab:Lattice_Parameters}
	\begin{tabular*}{\textwidth}{@{\extracolsep{\stretch{1}}}*{7}{r}@{}}
		\toprule
		\toprule
		Structure & a(\si{\angstrom}) & b(\si{\angstrom}) & c(\si{\angstrom}) & $\alpha$(\textdegree)& $\beta$(\textdegree)& $\gamma$(\textdegree) \\
		\midrule
		Bulk        & 6.41695      & 6.52047    & 7.28252 & 91.8128 & 103.5630 &  118.8390\\
		
		\\
		
		ML          & 6.41910      & 6.52306    & 45 & 90.7434 & 95.7909 &  118.8366\\
		
		\\
		
		BL          & 6.41910      & 6.52306    & 28.66062  & 84.0127  & 89.7412 &  61.1634\\	
		\bottomrule                             
	\end{tabular*}
	
\end{table*}


\begin{table}[!ht]
	\centering
	\setlength\tabcolsep{0pt} 
	\caption{Bandgap of bulk, BL and ML variants of $\text{ReS}_2$ at different levels of theory (with spin-orbit coupling). The gap increases from LDA to QS\emph{GW} level. The effect of screening is only moderately increased when two particle interactions are added (via a BSE, W $\rightarrow \hat{W}$) are added, thus only weakly decreasing the QS\emph{GW} band gap. }
	\label{tab:results_band_structure}
	\begin{tabular}{ p{2cm} p{2cm} p{2cm} p{2cm}}
		
		\multicolumn{4}{c}{\textbf{Band Gap(\si{\electronvolt})}}\\
		\toprule
		
		\textbf{Theory} & \textbf{LDA}  & \textbf{QS\emph{GW}} & \textbf{QSG}$\widehat{\textbf{W}}$ \\
		\toprule	
		\toprule	
		\textbf{Bulk}          & 1.15     & 1.75 & 1.7  \\
		
		\\
		
		\textbf{ML}          & 1.29      & 2.75 & 2.66 \\
		
		\\
		
		\textbf{BL}          & 1.23      &  2.35 &  2.3  \\
		
		\bottomrule
	\end{tabular}
\end{table}



\section{\textsc{Results and Discussions}}
\subsection{\textsc{Quasiparticle Energies and Band Structure}}\label{sec: band-str}

\begin{figure*}[htb]
	
	\subfloat[\label{bse-mono:a}]{%
		\includegraphics[scale=0.37]{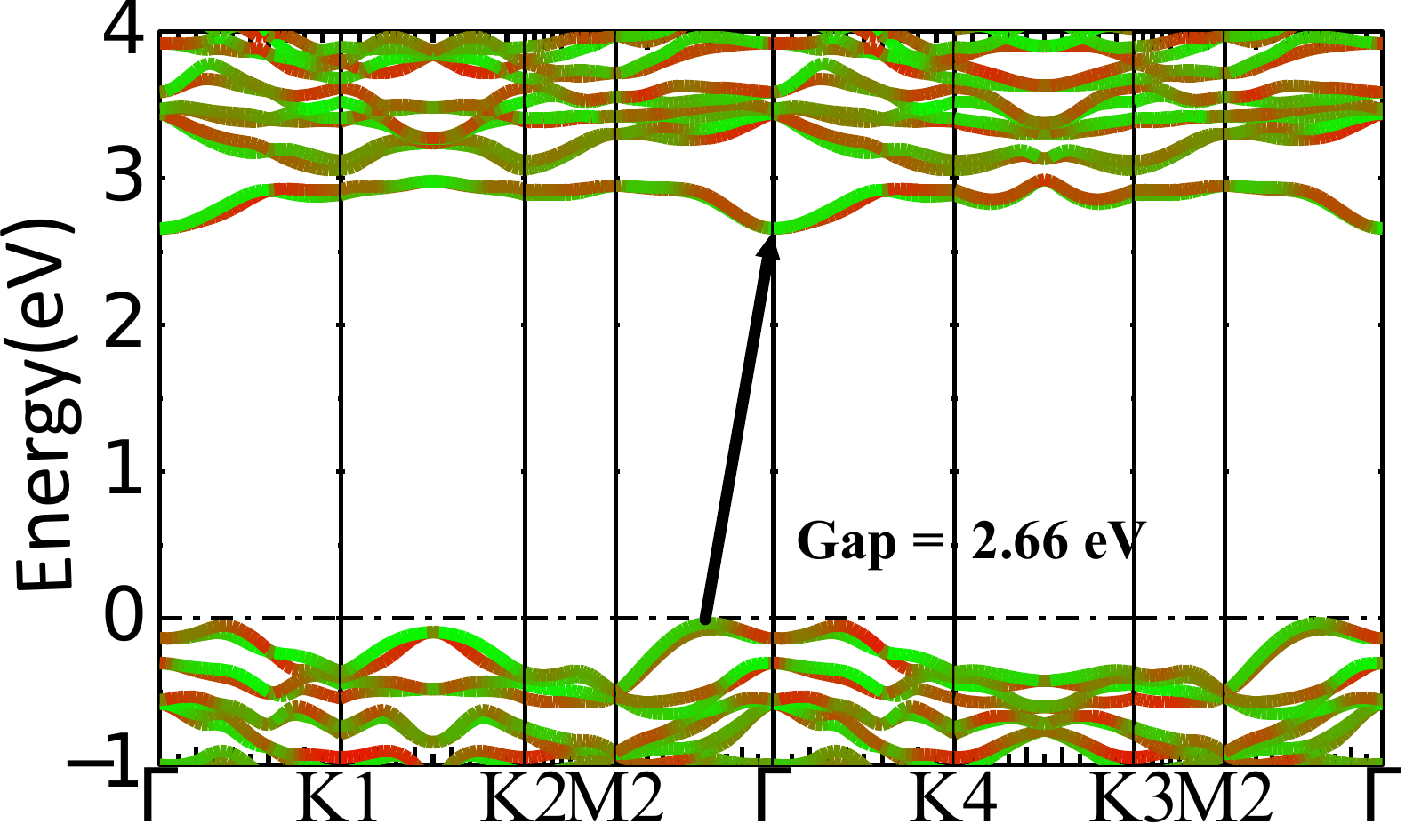}%
	}
	\subfloat[\label{bse-bi:b}]{%
		\includegraphics[scale=0.37]{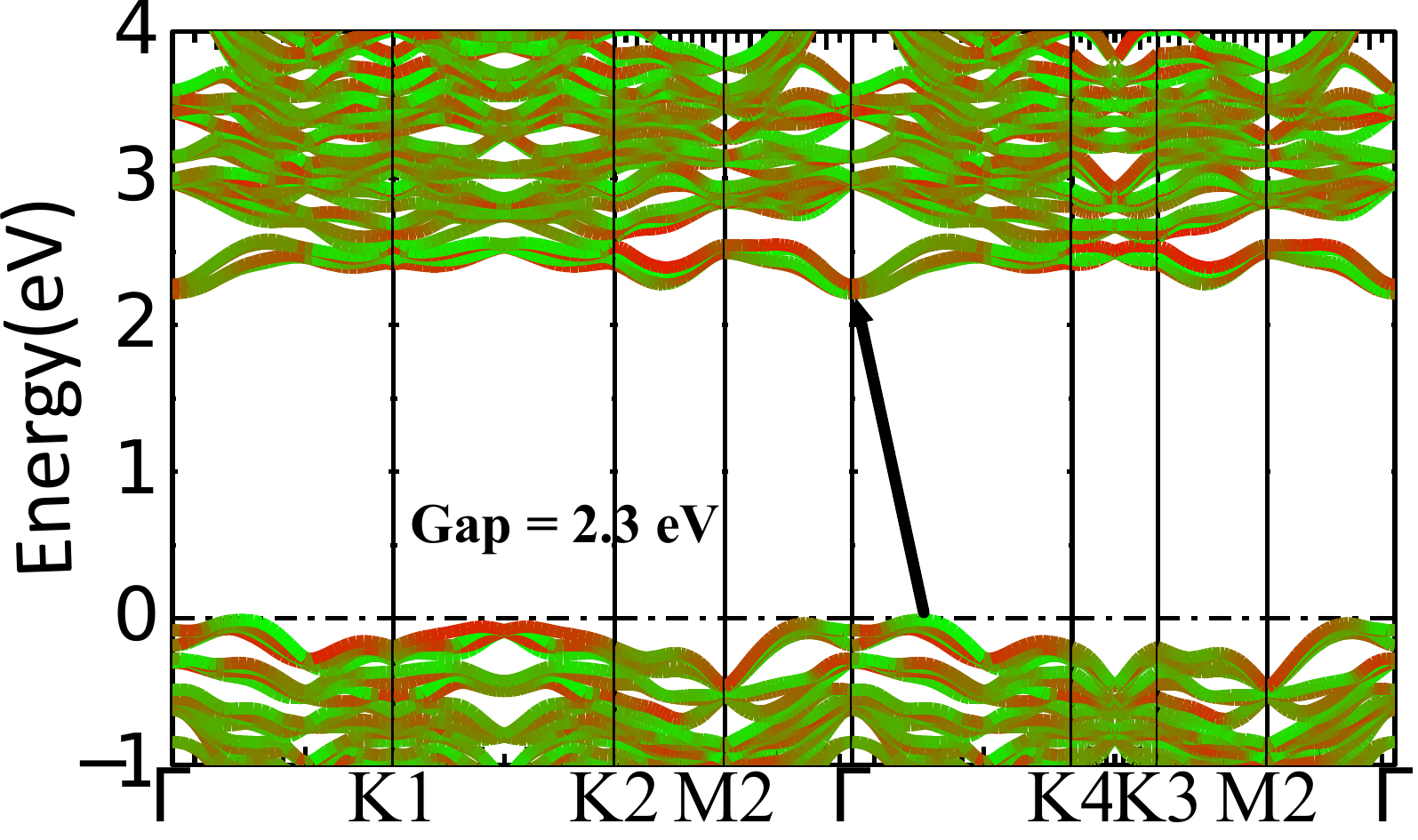}%
	}
	\subfloat[\label{bse-bulk:c}]{%
		\includegraphics[scale=0.37]{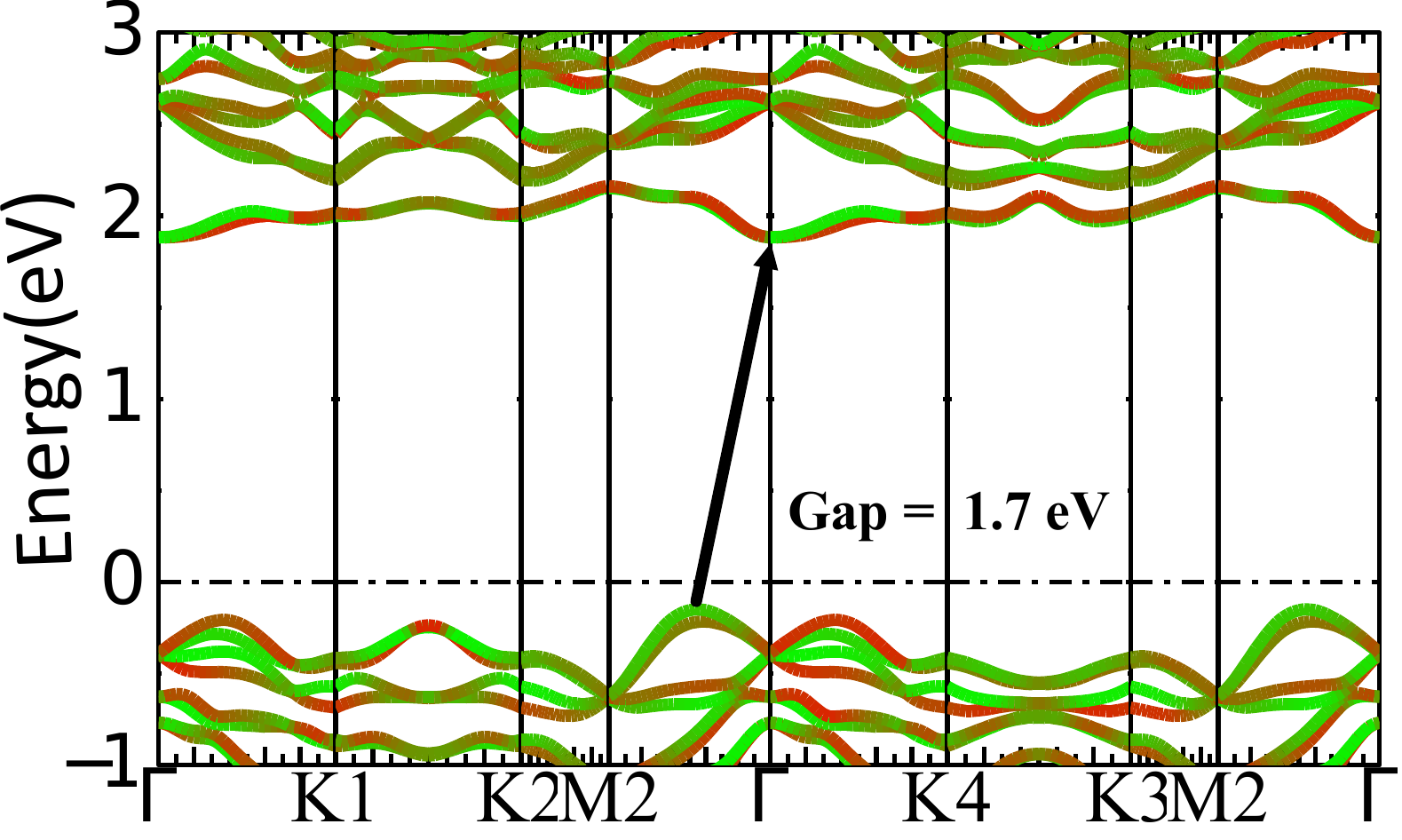}%
	}
	
	\caption{\bse band structures (with spin-orbit coupling) with contributions from Re (Red) and S (green). The nature of the band gap is indirect for all the variants of $\text{ReS}_2$ with values,  \SI{2.66}{\electronvolt} for ML (left), \SI{2.3}{\electronvolt} for BL (center) and \SI{1.7}{\electronvolt}  for bulk (right).}
	\label{fig:bse-bandgap}
	
\end{figure*}

\begin{table*}[!ht]
	\centering
	\setlength\tabcolsep{0pt} 
	\caption{Dielectric Constant(real part of the dielectric response at $\omega =0$) calculated using \bse. As the
          dimensionality of $\text{ReS}_2$ is lowered, the frequency range where Re($\epsilon_{xx}$) $\cdot$
          Re($\epsilon_{yy}$) is less than zero becomes smaller.}
	\label{tab:optical-spectra}
	\begin{tabular}{ p{1.5cm} p{1.5cm} p{1.5cm} p{1.5cm} p{3.5cm} p{2.5cm}}
		
		\multicolumn{4}{c}{\textbf{\thead{Dielectric Constant \\(\si{\electronvolt})}}}\\
		\toprule
		
	 & $\epsilon_{\infty}^{xx}$  & $\epsilon_{\infty}^{yy}$ & $\epsilon_{\infty}^{zz}$ & \textbf{\thead{Plasmonic frequency \\ range \\(\si{\electronvolt})}} & \textbf{\thead{Exciton \\  Binding energy \\(\si{\electronvolt}) }}\\
		\toprule	
		\toprule	
		\textbf{Bulk}          & 9.67     & 9.37 & 6.21 & 6.02 - 6.78 (0.76) & -  \\
		
		\\
		
		\textbf{BL}          & 6.97      &  7.09 &  2.66 & 6.65 - 7.08 (0.43) & 0.3 \\
		
		\\
		
		\textbf{ML}          & 2.97      & 3.19 & 1.42 & - & 0.74 \\
		
		\bottomrule
	\end{tabular}
\end{table*}

\begin{figure*}[htb]
	
	\subfloat[\label{mono-xx-yy-bse-real:a}]{%
		\includegraphics[scale=0.33]{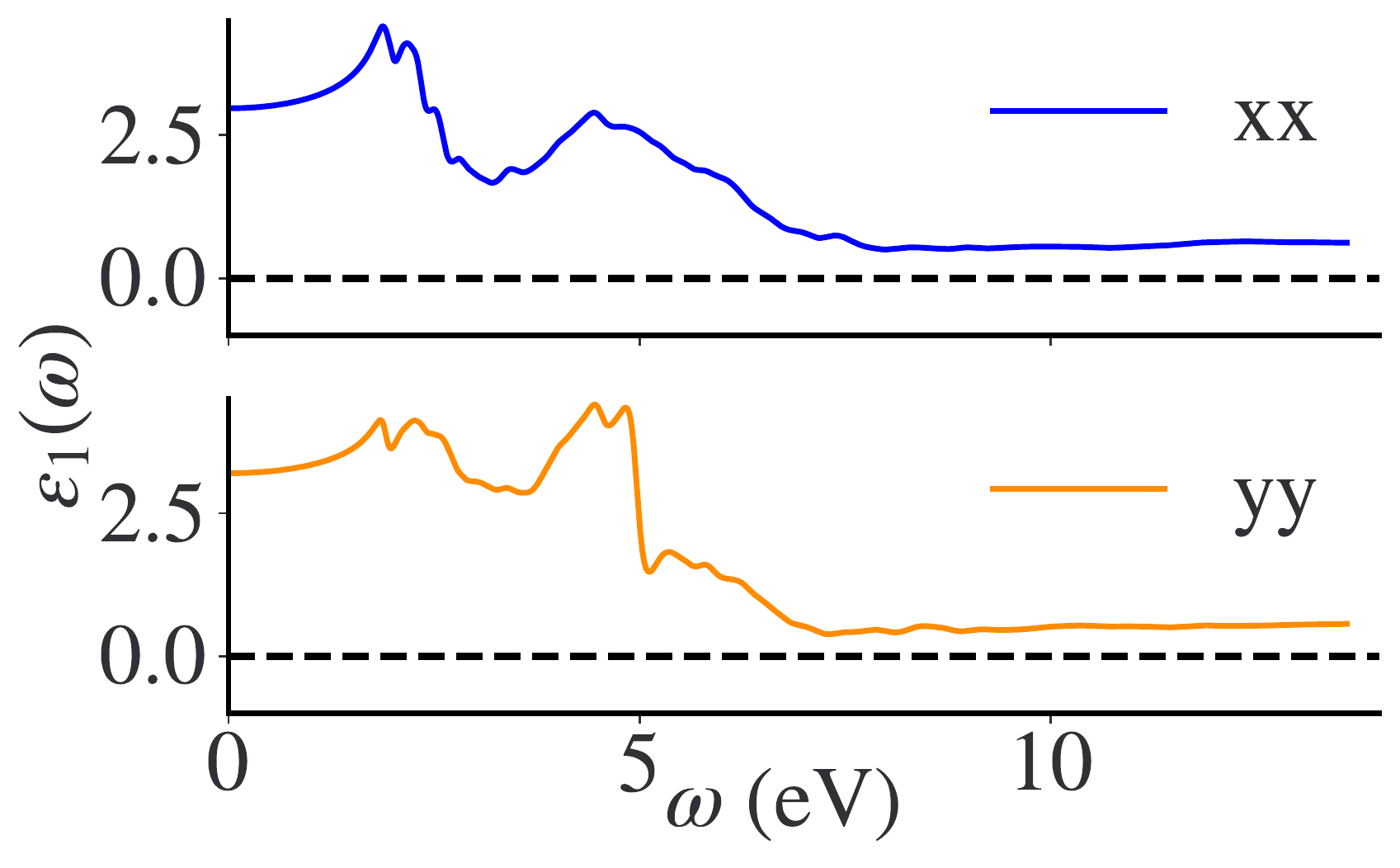}%
	}
	\subfloat[\label{bi-xx-yy-bse-real:b}]{%
		\includegraphics[scale=0.33]{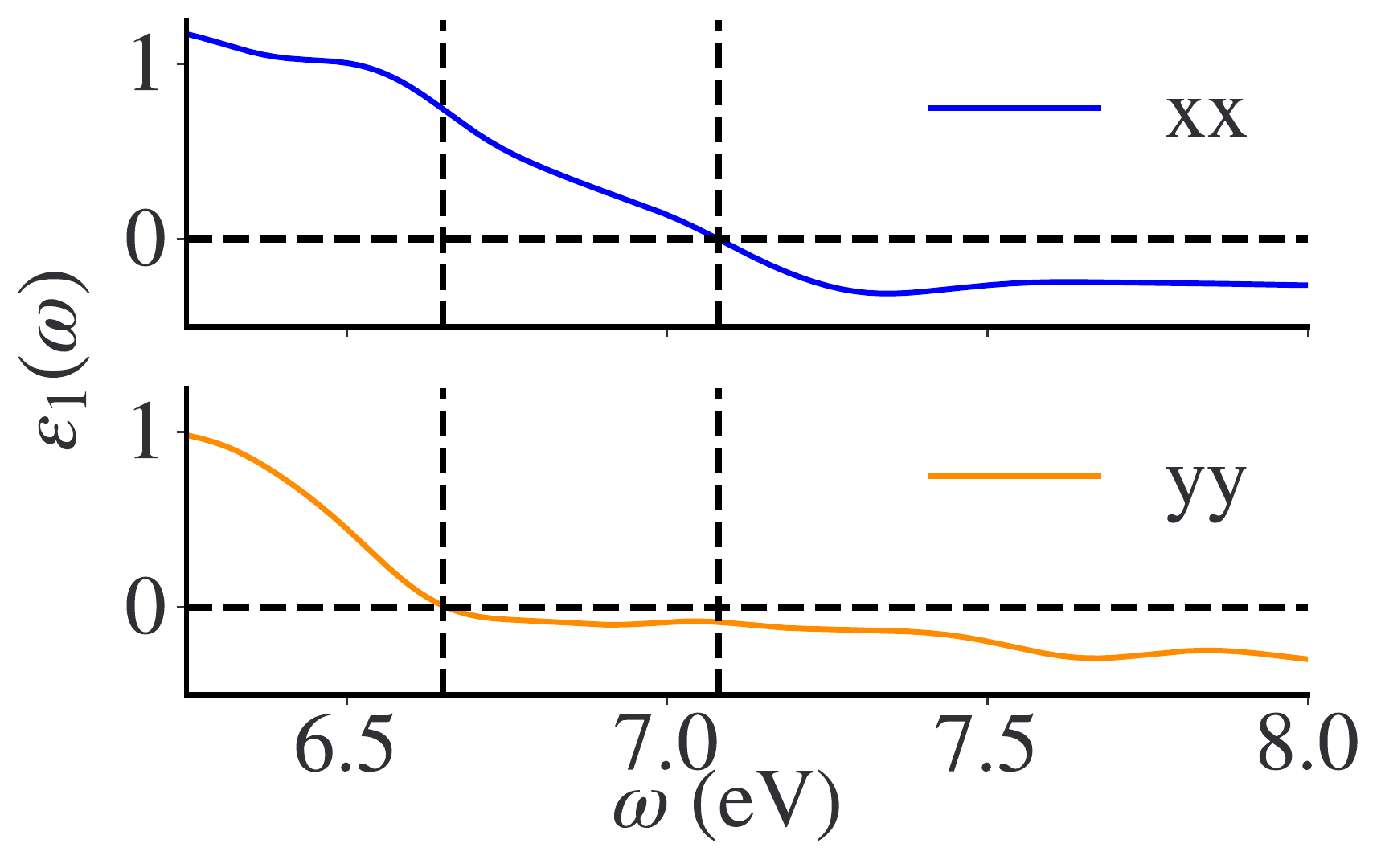}%
	}
	\subfloat[\label{bulk-xx-yy-bse-real:c}]{%
		\includegraphics[scale=0.33]{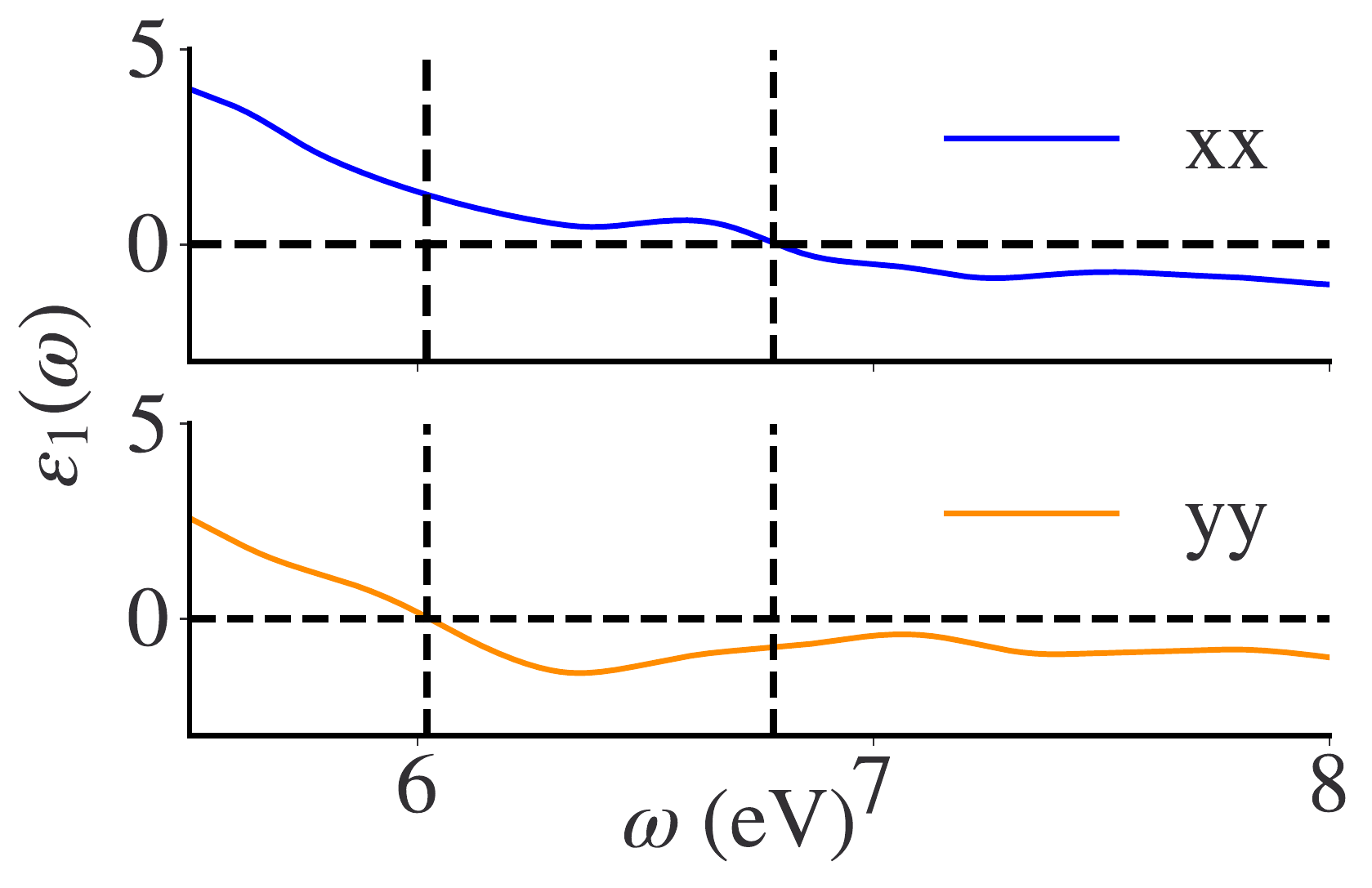}%
	}
	
	\subfloat[\label{mono-xx-yy-bse-img:d}]{%
		\includegraphics[scale=0.33]{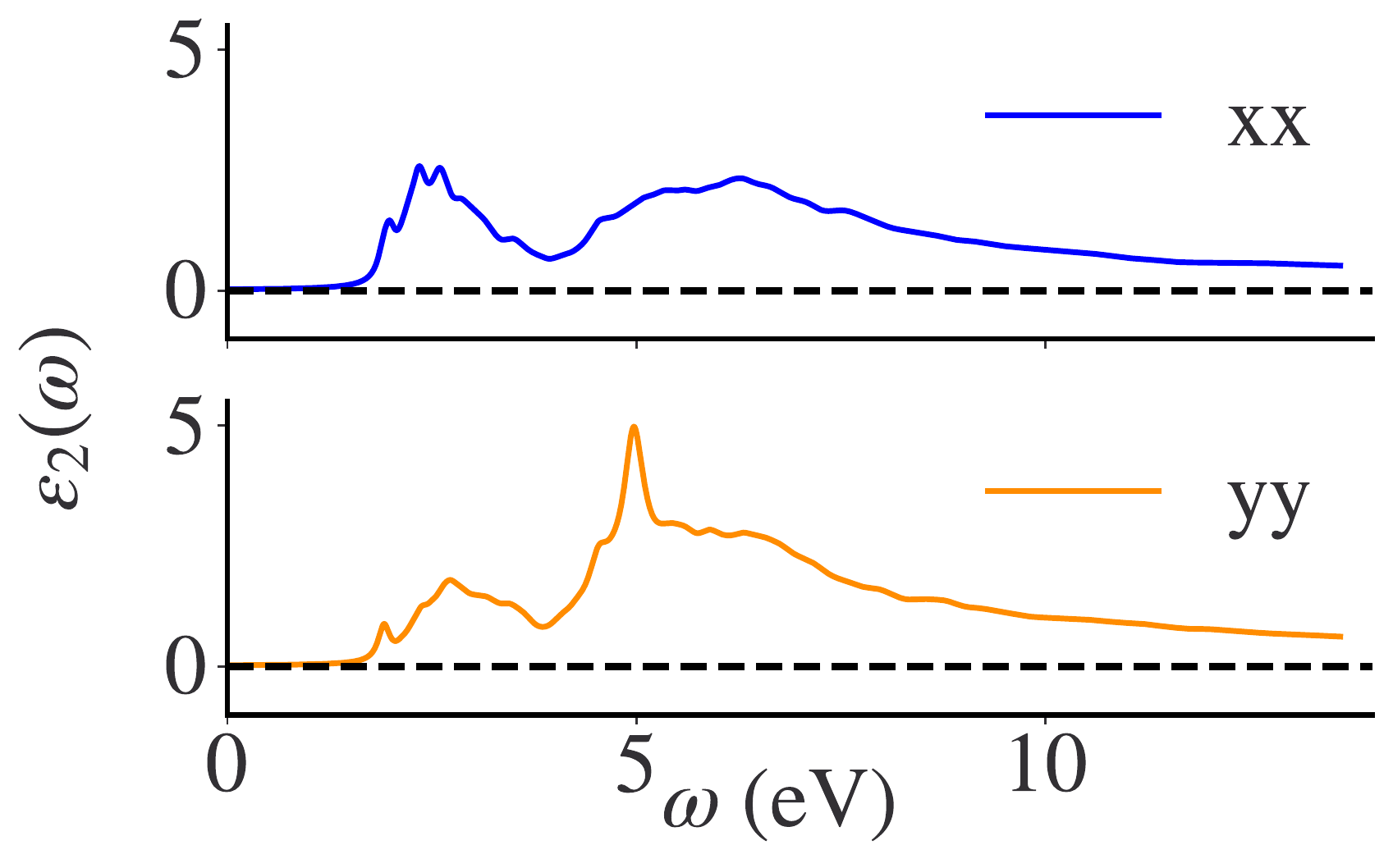}%
	}
	\subfloat[\label{bi-xx-yy-bse-img:e}]{%
		\includegraphics[scale=0.33]{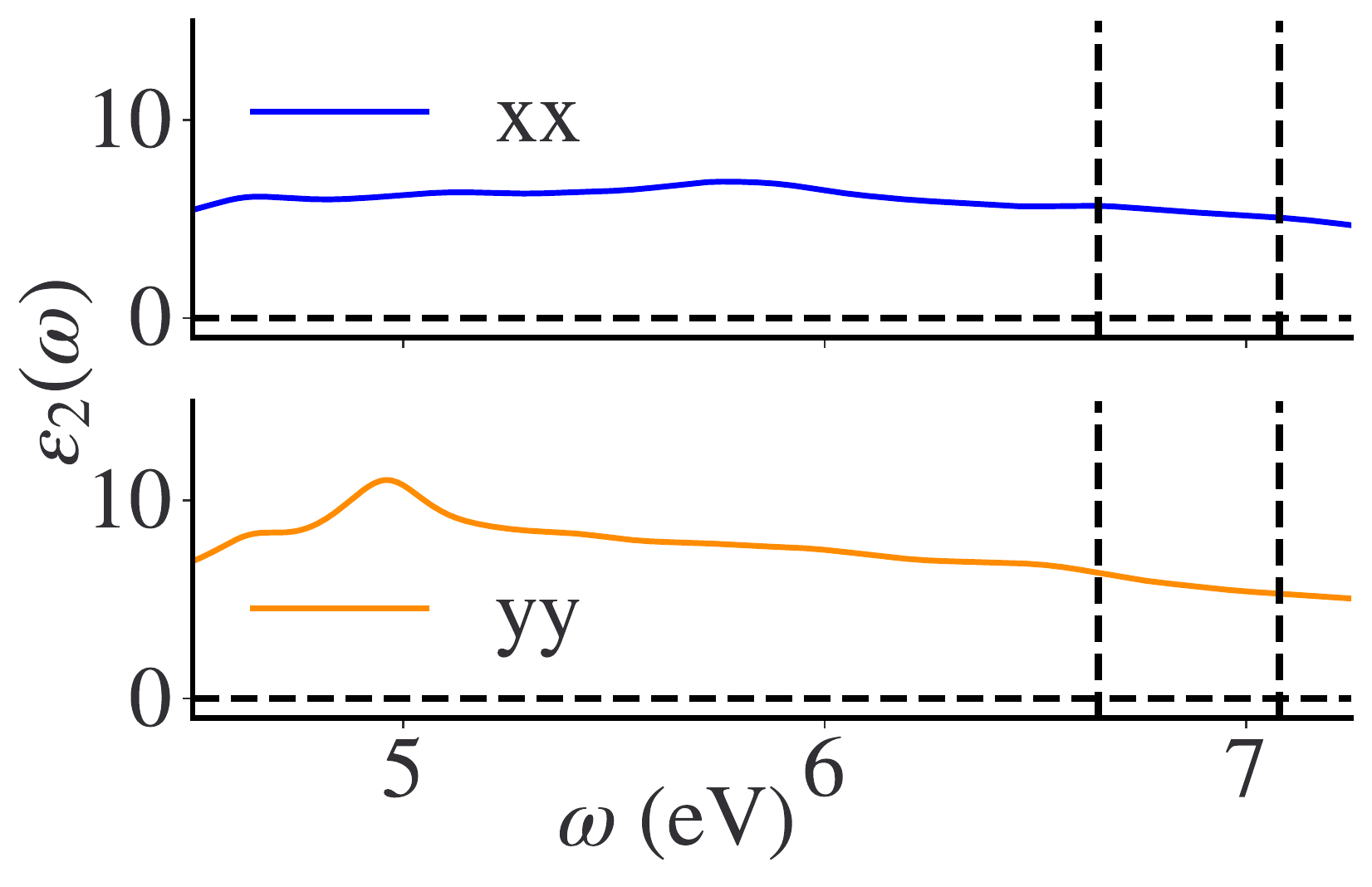}%
	}
	\subfloat[\label{bulk-xx-yy-bse-img:f}]{%
		\includegraphics[scale=0.33]{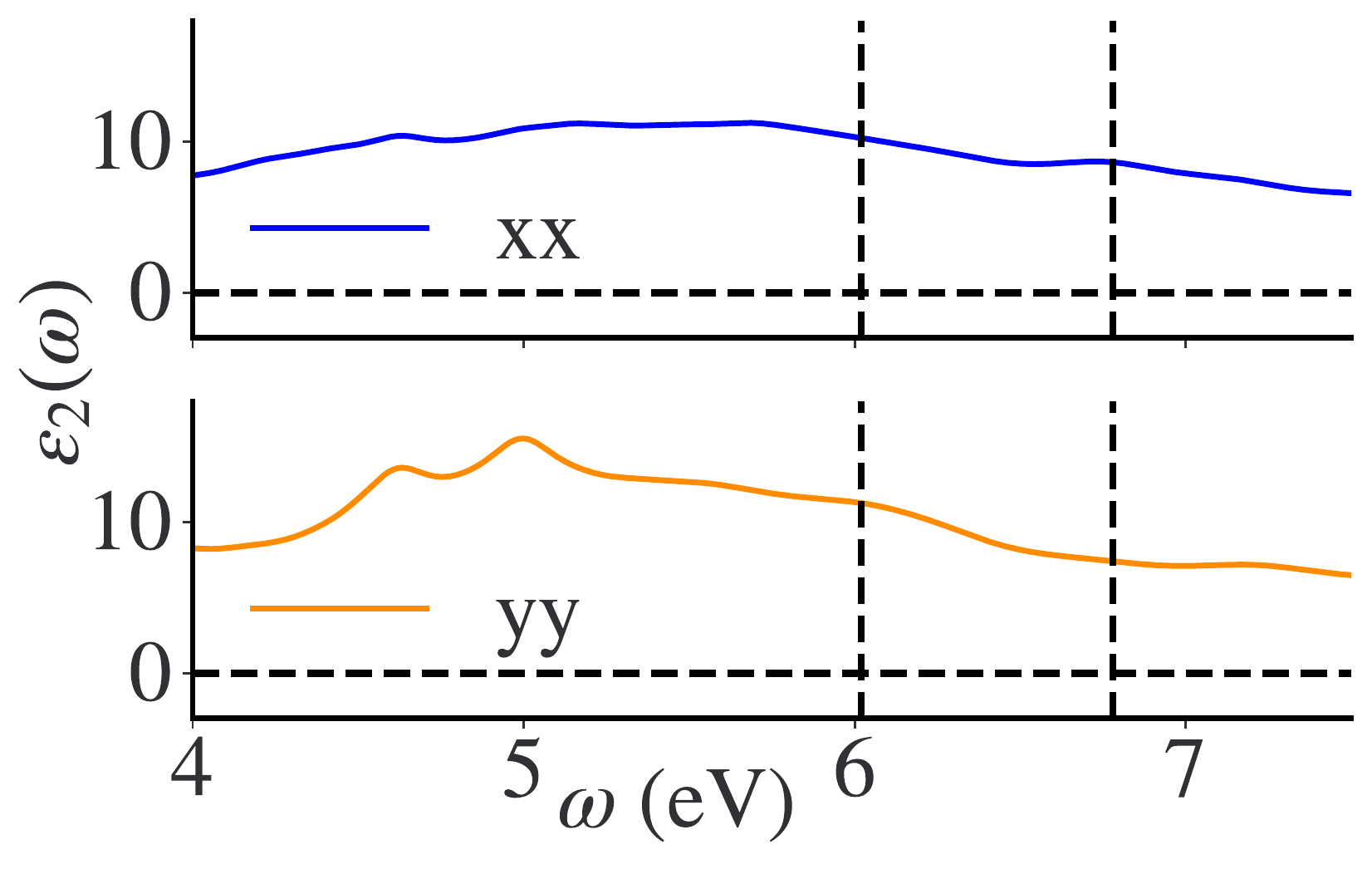}%
	}
	
	\caption{Real part($\epsilon_1$) of the dielectric response (top row) and imaginary part (bottom row) along $x$ and $y$ direction for ML (left), BL (center) and bulk (right). The vertical dashed line marks the plasmonic frequency range, where $\epsilon_1^{xx}(\omega) \times \epsilon_1^{yy}(\omega) < 0$.}
	\label{fig:bse-real-img}

\end{figure*}

The nature of the bandgap of $\text{ReS}_2$ has been widely debated in the literature. In a typical TMD family, the bandgap is direct when their thickness is reduced to the monolayer, ensuring that coupling with light is strong.  One study from 2014 \cite{tongay2014monolayer} reported direct bandgap for bulk ReS$_{2}$, thus generating considerable interest in the system. However, both older studies such as \cite{ho1997optical,ho1997temperature} and more recent studies such as \cite{aslan2016linearly, gutierrez2016electroluminescence} report that the bulk $\text{ReS}_2$ is an indirect-bandgap semiconductor. 
	
The electronic band structures (with spin-orbit coupling included) for bulk, BL and ML are shown in
Fig. \ref{fig:bse-bandgap} and the band gaps at different levels of theory are summarized in Table
\ref{tab:results_band_structure}. The free-standing ML of $\text{ReS}_2$ has been simulated for these calculations with
the parameters shown in Table \ref{tab:Lattice_Parameters}. Similar to prior work on monolayers of chromium trihalides
\cite{acharya2021electronic}, we check for convergence and scaling of band gap and the dielectric constant
$\epsilon_{\infty}$ with vacuum size. We obtain the LDA band gap of $\sim$\SI{1.29}{\electronvolt} which is
significantly lower than the QS\emph{GW} band gap of $\sim$\SI{2.75}{\electronvolt}. LDA is known to underestimate the band gaps in semiconductors, and the enhancement in
the QS\emph{GW} band gap relative to LDA is standard~\cite{van2006quasiparticle}.  QS\emph{GW} usually
  overcorrects the gap because \emph{W} is universally too large within the random phase approximation (RPA),
and for the same reason it underestimates the dielectric constant $\epsilon_{\infty}$
  \cite{cunningham2018effect}.  Adding ladders largely eliminates both tendencies.  In the present case extending
$\text{QS}GW{\rightarrow}\text{QS}G\widehat{W}$ causes only a modest reduction in the gap, to
$\sim$\SI{2.66}{\electronvolt}, suggesting insignificant corrections to the self-energy originating from BSE. This
result is consistent with previous theoretical work \cite{zhong2015quasiparticle} on ML-$\text{ReS}_2$. The self-energy
and the reduced screening increase the band gap and modify the band topology, which is observed in the
ML-$\text{ReS}_2$. The band gap at the level of LDA is direct, but the two-particle interactions lower the valence band
maxima (VBM), which was at the $\Gamma$ point by about \SI{150}{\milli \electronvolt}. A similar kind of change in band
topology has been observed for $\text{ReSe}_2$ in \cite{zhong2015quasiparticle} but is absent in other some theoretical
work \cite{echeverry2018theoretical}.
	
We obtain a similar band topology in bulk and BL variants of $\text{ReS}_2$; however, the band gap values are
$\sim$\SI{1.7}{\electronvolt} and $\sim$\SI{2.3}{\electronvolt} respectively. The nature of band gap in $\text{ReS}_2$
is different from more commonly studied semiconducting TMDs (e.g., MoS$_{2}$, WS$_{2}$, etc.), where the bulk and
few-layer variants show indirect band gap and ML is direct. In this work, we observe a direct band gap at the LDA level
and an indirect band gap for all the variants at the \bse level. This nature of the quasiparticle band gap does not
conflict with experimental measurement because, different from our calculated free-standing cases, these measured
samples are on substrates and are inevitably doped. The resulting self-energy corrections will be reduced under these
conditions, resulting in a slightly indirect band gap for the measured samples. We note that the direct to indirect
transition may not be sharp because the energy difference between the direct and indirect gap is small and external
perturbations can affect the conclusion. However, our observations on the indirect nature of the band gap is important for many reasons. In most monolayer TMDs, the gap is direct and it makes them sufficiently bright and also the electron-hole radiative lifetimes are extremely small (often in picoseconds). While in systems with indirect band gaps, radiative and non-radiative processes compete, since the indirect states have longer lifetimes making them candidates for optoelectronics and photovoltaics~\cite{pnas}. 


\subsection{\textsc{Optical Absorption Spectra : Hyperbolic Plasmons}}\label{sec: optical-spectra}

\begin{figure*}[htb]
	
	\subfloat[\label{mono-plasmon:a}]{%
		\includegraphics[scale=0.33]{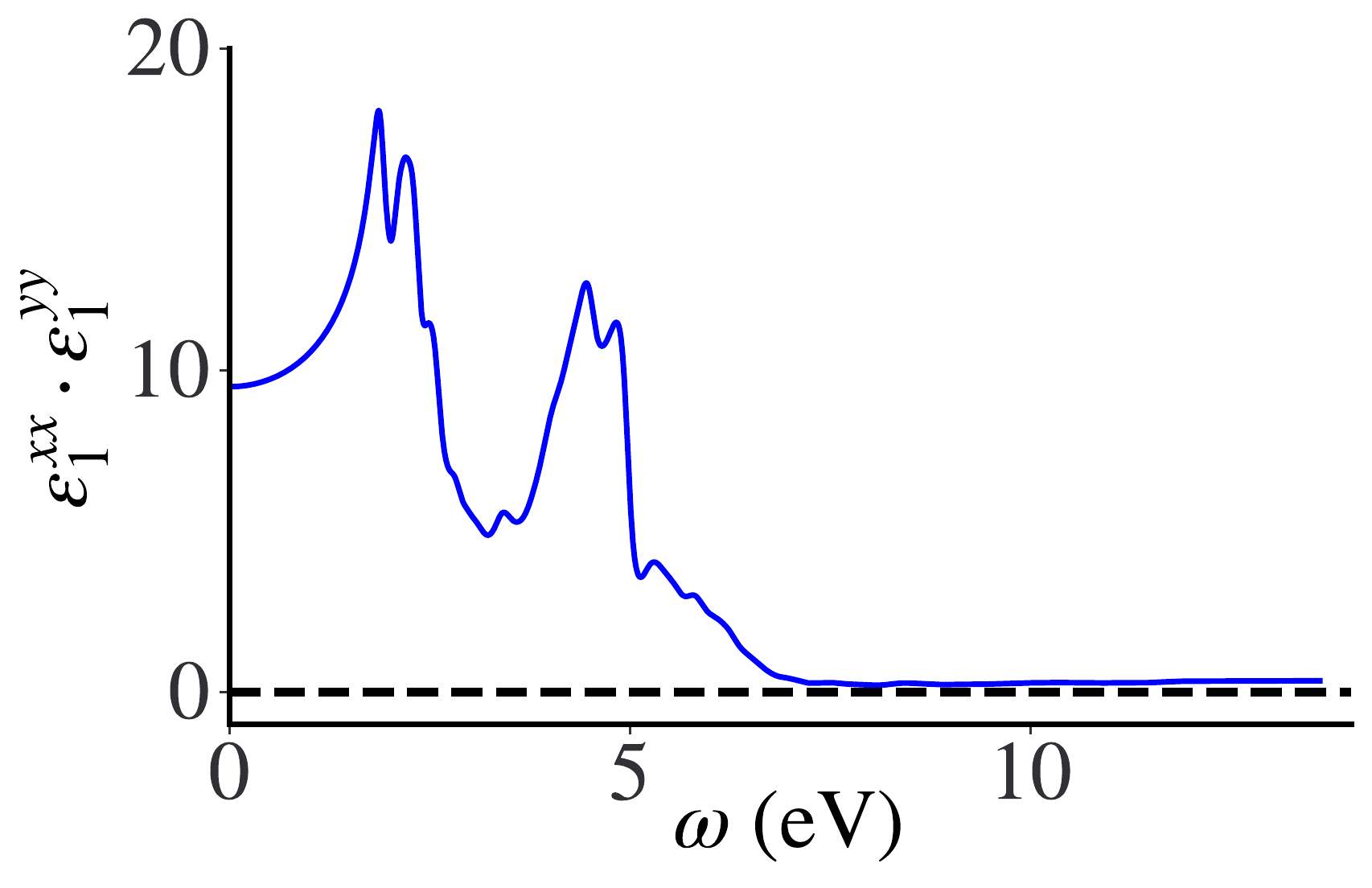}%
	}
	\subfloat[\label{bi-plasmon:b}]{%
		\includegraphics[scale=0.33]{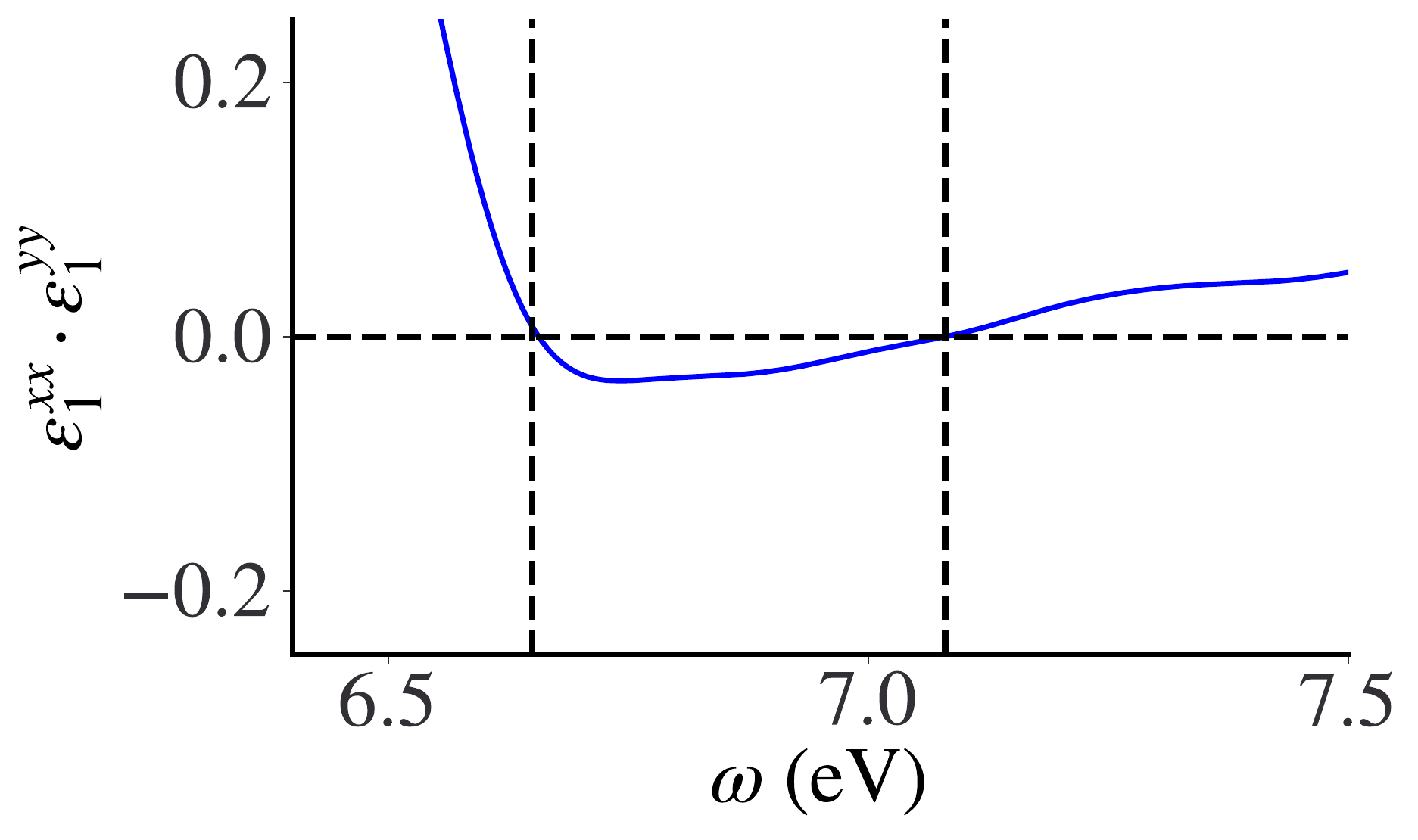}%
	}
	\subfloat[\label{bulk-plasmon:c}]{%
		\includegraphics[scale=0.33]{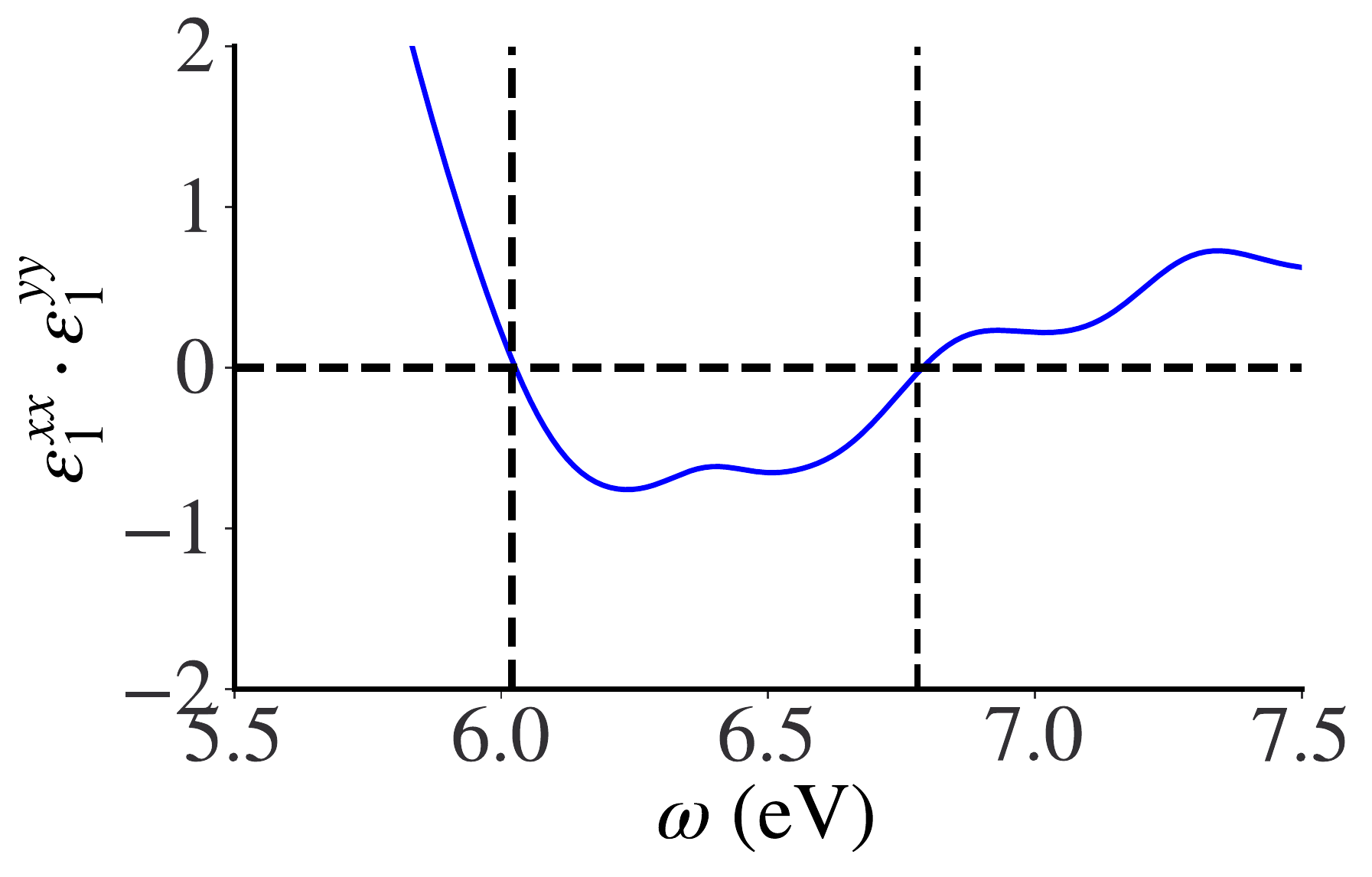}%
	}
	
	\caption{The product of the real part of the dielectric response along X($\epsilon_1^{xx}(\omega)$) and Y($\epsilon_1^{yy}(\omega)$) direction for ML(left),BL(center) and Bulk(right). The vertical dashed line marks the frequency range in which the product is negative. This frequency range is hyperbolic plasmonic frequency range.}
	\label{fig:plasmons}
	
\end{figure*}

The anisotropic optical absorption has been previously studied using DFT \cite{ho2004optical}. Experimentally
\cite{aslan2016linearly} it is demonstrated that the reduced crystal symmetry of $\text{ReS}_2$ leads to anisotropic
optical properties that persist from the bulk down to the monolayer limit. The absence of excitonic correlations and
underestimated band gaps in LDA studies hide several physical consequences in $\text{ReS}_2$. Advanced theoretical
studies such as \cite{zhong2015quasiparticle} tackle anisotropic optical responses at the BSE level for monolayer
$\text{ReS}_2$, where ladder vertex corrected optical properties are computed on top of a single shot DFT based
G$_{0}$W$_{0}$ one-particle description.  
	
The large structural anisotropy in 2D materials, for example, a 4:3 anisotropy of the in-plane lattice constants in black
phosphorus~\cite{nemilentsau2016anisotropic} and solid
nitrogen~\cite{rudenko2022electronic,ji2020nitrogen,PhysRevLett.124.216001}) makes them perfect candidates for
hyperbolic materials and a natural place to look for HP. This offers new possibilities as hyperbolic materials showcase a wide variety of interesting properties, such as modes which transport heat by photon tunnelling with a high efficiency close to the theoretical limit\cite{biehs2012hyperbolic}, and broadband absorption \cite{riley2017near}. We first define the condition for hyperbolicity. The hyperbolic region appears when
\begin{equation}
		\epsilon_1^{xx}(\omega) \cdot \epsilon_1^{yy}(\omega) < 0 
\end{equation} 
where, $\epsilon_1^{xx}(\omega)$ and $\epsilon_1^{yy}(\omega)$ are the real part of the dielectric response along the x and y direction respectively. We assume that $\epsilon_1^{xy}(\omega)=0$ by symmetry and, thus, $x$ and $y$ are the principal directions of the dielectric permeattivity. For different variants of $\text{ReS}_2$, the real and imaginary part of dielectric response is plotted in Fig. \ref{fig:bse-real-img}. We observe that significant difference in optical response for incident polarized light along different directions. While ML-$\text{ReS}_2$ (Fig. \ref{fig:bse-real-img}, left panel), hosts some strongly bound anisotropic excitons deep inside the one-particle gap, HP are absent. For the BL-$\text{ReS}_2$ (Fig. \ref{fig:bse-real-img}, center panel), the Re($\epsilon_{yy}$) goes to negative at $\omega= \SI{6.65}{\electronvolt}$, which results in a hyperbolic region starting at that frequency with an energy window of \SI{0.43}{\electronvolt}. For bulk-$\text{ReS}_2$ this energy window increases to \SI{0.76}{\electronvolt}. The sign change is key to the appearance of the hyperbolic region and it becomes more apparent in Fig. \ref{fig:plasmons} where we plot the product $\epsilon_1^{xx}(\omega) \times \epsilon_1^{yy}(\omega)$ which becomes negative in the energy window.  $\epsilon_2$ remains large for both bulk and BL in the hyperbolic energy window suggesting large damping of the HP. In the BL, these plasmons are less damped compared to bulk. Also, note that these HP in ReS$_{2}$ are in ultraviolet range, in contrast to the infrared HP in CuS nanocrystals~\cite{cordova2019anisotropic}.

The inherent anisotropy in ReS$_{2}$ provides an opportunity to tune its magnitude by applying strain. We apply uni-directional strains ($\gamma$) along $x$ and $y$ respectively and explore the hyperbolic region. We apply up to 4\% strain and see that the ML never hosts HP. However, in bulk  on application of compressive unidirectional strain ($\gamma_{x}$) along $x$, the HP window increases upto $\sim$1.3 eV. This enhanced HPs also have lesser damping compared to the un-strained compound. On the other hand, $\gamma_{y}$ reduces the HP window. In the BL, application of $\gamma_{x}$, almost entirely kills the HP window while $\gamma_{y}$ enhances the HP window and also hosts less damped plasmonic modes. In short, we observe that while strain can be used to tune the hyperbolic energy window and the lifetimes of the plasmons, we could not produce HPs in the monolayer sample under any condition. However, the HPs remain pretty robust in both the bulk and BL variants, also their stability could be enhanced on selective applications of uni-directional strain.

\section{Conclusion} \label{sec: conclusion}

Anisotropy is a key to tuning material properties. Discontinuities at surfaces, residual strains and metamaterials
have been used as platforms for realizing anisotropic optical properties. Naturally occurring structurally anisotropic materials are not necessarily hyperbolic always.

In this work we show that the structural anisotropy in $\text{ReS}_2$, even though much weaker compared to materials like black Phosphorus and solid Nitrogen, leads to the occurrence of hyperbolic plasmons in a narrow energy window. The plasmonic resonances can be tuned by controlling the number of layers in the far ultraviolet frequency range. The ability of such a tunability of the plasmons opens up new opportunities in regard to optoelectronic devices. We further show that the hyperbolic region and its stability can be enhanced by unidirectional strains. 

\section{Acknowledgement} \label{sec: Acknowledgement}
MIK and SA are supported by the ERC Synergy Grant, project 854843 FASTCORR (Ultrafast dynamics of correlated
electrons in solids). MvS and DP were supported by the Computational Chemical Sciences program
within the Office of Basic Energy Sciences, U.S. DOE under Contract
No. DE-AC36-08GO28308. This research used resources of the National
Energy Research Scientific Computing Center (NERSC), award
BES-ERCAP0021783, under DOE Contract No. DE-AC02-05CH11231. We acknowledge PRACE
for awarding us access to Irene-Rome hosted by TGCC, France and Juwels Booster and Cluster, Germany. 
	
\bibliography{references}
\end{document}